\documentstyle[12pt]{article}
\def\lb{\label}
\def\be{\begin{equation}}
\def\ee{\end{equation}}
\def\ba{\begin{eqnarray}}
\def\ea{\end{eqnarray}}
\def\ds{\displaystyle}

\def\e{{\rm e}}
\def\vr{{\bf r}}
\def\vk{{\bf k}}

\def\du{{\dot u}}
\def\ddu{{\ddot u}}

\begin{document}

\title{
   \begin{flushright} \begin{small}
     DTP-MSU/01-22 \\
  \end{small} \end{flushright}
%\vspace{.5cm}
%%%%  Title %%%%
{\bf Radiation reaction in  various dimensions } }
%%%%%  Authors  %%%%
\author{
   D.V. Gal'tsov
\thanks{Supported by RBFR. Email:  galtsov@grg.phys.msu.su} \\
{\it Department of Theoretical Physics,}\\
       {\it Moscow State University, 119899, Moscow, Russia}
}
\maketitle
\begin{abstract}
We  discuss the radiation reaction problem  for an electric
charge moving in flat space-time of arbitrary dimensions. It is
shown that four is the unique dimension where a local differential
equation exists  accounting for the radiation reaction and
admitting a consistent mass-renormalization (the Dirac-Lorentz
equation). In odd  dimensions the Huygens principle does not
hold; as a result, the radiation reaction force depends on the
whole past history of a charge (radiative tail). We show that the
divergence in the tail integral can be removed by the mass
renormalization only in the 2+1 theory. In even dimensions higher
than four, divergences can not be  removed by a renormalization.

\bigskip
PACS no: 04.20.Jb, 04.50.+h, 46.70.Hg
\end{abstract}

\section{Introduction}
In this note we address the question whether a generalization of
the Dirac-Lorentz equation for a radiating charge in classical
electrodynamics exists in space-times of dimension other than
four. Apart from a purely academic interest, one is naturally led
to higher-dimensional radiation problems in the brane-world set
up. Although in the latter case it is not the {\it vector} field
radiation which is
  interesting to study (vector fields do not live in the
bulk), we believe  that basic dimensionality conditioned features
of  radiation reaction  are similar for  any spin.

If one tries to generalize the Dirac's derivation of the
radiation reaction force \cite{Di38} to arbitrary (flat)
space-time dimensions, two kinds of obstacles can be foreseen.
First, the divergence of the proper  Coulomb potential of a charge
in higher dimensions is stronger than in four, so the divergence
of the self-force acting on it will be stronger. Another unusual
feature is encountered in space-times of {\it odd} dimensions. It
is well-known that the Huygens principle does not hold in odd
dimensions, and radiation develops a tail, similar to that known
in the four-dimensional {\em curved} space-time
\cite{DWBr60,Ho68}. Therefore, it can be expected that the
dynamics of a radiating charge in odd dimensions will be governed
by an integro-differential equation.

We discuss  the difference between   Green functions of the
scalar D'Alembert equation in neighboring dimensions in  detail
and suggest a simple physical interpretation of the Huygens
principle violation  in odd dimensions.  The Green function in odd
dimensions can be obtained by integrating the neighboring higher
even-dimensional Green function  over the extra dimension, this
is equivalent to take the source smeared along an extra
coordinate. Radiation, coming to a given point from distant parts
of the line source, produces a tail. Passing to the next lower
dimension, one finds that tails are canceled by a destructive
interference,  when summing up along  the second extra dimension.

The case of the radiating charge in $2+1$ (minimal) Maxwell
theory is shown to be consistent, and the integral version of the
Dirac-Lorentz equation is derived. However all higher-dimensional
generalizations, both in even and odd dimensions, fail because of
impossibility of absorbing divergences by  a renormalization of
parameters.

We use the 'mostly minus' metric signature.
%%%%%%%%%%%%%%%%%%%%%%%%%%%%%%%%%%%%%%%%%%%
\setcounter{equation}{0}
\section{General setting}
For reader's convenience we start by recalling briefly the
original Dirac's derivation \cite{Di38}. Consider the charge
equation of motion with a 'bare' mass $m_0$ \be\lb{bareeq} m_0
\du^\mu=e {F^\mu}_\nu u^\nu, \ee where $u^\mu=dx^\mu\ds$ is a
tangent vector to the world-line $x^\mu=x^\mu(s)$, and   the field
strength  denotes the sum $F=F^{ext}+F^{ret}$  of an external
field and the (retarded) proper field of the charge. The potential
of the latter ($F=dA$) is a solution of the D'Alembert equation
\be\lb{dal} \Box A^\mu(x)=-4\pi e \int u^\mu(s)
\delta^4(x-x(s))ds. \ee
 Omitting for the moment the external
field, we adopt the standard decomposition of the proper  field of
the  charge $A^{ret}=A^{self}+A^{rad}$ : \be \lb{Asr}
A^{self}=\frac12 (A^{ret}+A^{adv}),\quad A^{rad}=\frac12
(A^{ret}-A^{adv}). \ee

The retarded and advanced Green functions, satisfying the
equation
 \be\lb{dalg} \Box G^{ret,adv}(x-x')=-4\pi
\delta^4(x-x'), \ee in four dimensions are given by \be\lb{G4}
G^{ret,adv}(X)=\frac{\delta(T\mp R)}{R}, \ee where
$X^\mu=x^\mu-x'^\mu,\; T=t-t',\; R=|\vr-\vr'|$. Their
combinations corresponding to  (\ref{Asr}) are \be \lb{Gsr}
G^{self}=\delta{(T^2-R^2)},\quad
G^{rad}=\frac{T}{|T|}\delta{(T^2-R^2)}. \ee

Taking the value of the  electromagnetic field of the  charge on
its world-line leads to the following integral on the right hand
side of (\ref{bareeq}) \be f^\mu(s)=4
e^2\int\;X^{[\mu}(s,s')u^{\nu]}(s')u_\nu(s)\frac{d}{d X^2} G(X)
ds', \ee where $X^\mu(s,s')=x^\mu(s)-x^\mu(s'),\; X^2=X^\mu
X_\mu=T^2-R^2$. Due to the presence of delta-functions in both
Green functions $G^{self}$ and $G^{rad}$, only a finite number of
Taylor expansion terms in $\sigma=s-s'$ contribute to the
integral, since $X^2=\sigma^2+O(\sigma^4)$. In the
four-dimensional case it is sufficient to retain terms up to
$\sigma^3$: \be\lb{exp}
4X^{[\mu}(s,s')u^{\nu]}(s')u_\nu(s)=\du^\mu\sigma^2-
\frac23(\ddu^\mu+u^\mu\du^2 )\sigma^3 +O(\sigma^4). \ee The
leading terms in the expansions of derivatives of the Green
functions are \be\lb{sings} \frac{d}{d X^2}
G^{self}(X)=\frac{d}{d \sigma^2}\delta(\sigma^2),\qquad
\frac{d}{d X^2} G^{rad}(X)= \frac{d}{d
\sigma^2}\left(\frac{\sigma}{|\sigma|}\delta(\sigma^2)\right). \ee
Consequently, one encounters the following integrals for the
self-force \be A_l=\int_{-\infty}^\infty \sigma^l\,\frac{d}{d
\sigma^2}\delta(\sigma^2)\;d\sigma, \ee and for the radiation
reaction force \be B_l=\int_{-\infty}^\infty \sigma^l\,\frac{d}{d
\sigma^2}\left(
\frac{\sigma}{|\sigma|}\delta(\sigma^2)\right)\,d\sigma, \ee with
$l\geq 2$. The first integral is divergent for $l=2$,  equal to
zero for $l=3$ by parity, and vanishes for all $l>3$. The second
is zero for $l=2$ by parity and vanishes for all $l>3$. The
integral $B_3$ is finite, and, in order to disentangle the
integrand, it is convenient to regularize the expression by the
shift  of the argument $\sigma^2-\epsilon^2$ of the
delta-function, taking the limit $\epsilon=0$ at the end. One
finds $B_3=-1$. The divergent term $A_2$ enters in the equation
of motion multiplied by  ${\dot u}^\mu$, so it can be absorbed by
the renormalization of the mass, \be m_0-A_2=m, \ee and we obtain
the Dirac-Lorentz equation \be m \du^\mu=e {F^\mu}_\nu
u^\nu+\frac23(\ddu^\mu+u^\mu\du^2 ), \ee where  the external
force term is reintroduced.

This setting remains qualitatively unchanged when we pass to
space-times of arbitrary dimensions. But the singularity
structure of the Green functions will be different, so in
dimensions higher than four we will get larger number of
divergent integrals both for the self-force and the radiation
reaction force. Moreover, in space-time of odd dimensions the
Green functions do not contain delta-functions at all, therefore
instead of the differential equation (obtained in the case when
only a finite number of powers of $\sigma$ give a non-zero
contribution) we will be left with an integral equation.

%%%%%%%%%%%%%%%%%%%%%%%%%%%%%%%%%%%%%%%%%%%%%%%%%%%%%%%%%%%%%%%%%%%%%%%%%
\setcounter{equation}{0}
\section{Green functions in even and odd dimensions}
The retarded Green functions of the D'Alembert equation in any
dimensions can be presented in a unique way in the momentum
representation \be G_{n+1}^{ret}(t-t',
\vr-\vr')=\frac{2}{(2\pi)^n} \int \frac{\e^{-i\omega (t-t')+i\vk
(\vr-\vr')}\;d\omega d\vk} {(\omega-i\varepsilon)^2-\vk^2}, \ee
where $\vr,\, \vk$ are $n$-dimensional vectors. However, in the
coordinate representation, which  describes the waves propagation
more directly, they turn out to be qualitatively different in odd
and even dimensions. Performing  a contour integration over
$\omega$ and then integrating over the $n-1$-dimensional sphere
in the $\vk$-space one is led to the following representation
\cite{IvSo40,IvSo48} \footnote{The author is grateful to E.Yu.
Melkumova for these references.}: \be\lb{sonin}
G_{n+1}^{ret}\,=\,\frac{2\theta(T)}{(2\pi
R)^{(n/2-1)}}\,\int_0^\infty \;k^{n/2-1}\; \sin kT \;
J_{n/2-1}(kR)\; dk. \ee Here $T=t-t',\; R=|\vr-\vr'|$ and
$J_{n/2-1}$ is the Bessel function. Mathematically, the
distinction between odd and even dimensions is due to the fact
that for odd $n=2\nu+1$ (even dimension of space-time) the index
of the Bessel function is semi-integer, and this function is
expressible in terms of elementary functions: \be
J_{\nu-1/2}(kR)=\sqrt{\frac{2}{\pi}}(-1)^{\nu-1}\left(\frac{R}{k}\right)^{\nu-1/2}
\frac{d^{\nu-1}}{(RdR)^{\nu-1}} \frac{\sin kR}{R}. \ee For
$\nu=1$ one finds the Green function (\ref{G4}) which is
localized on the light cone surface. For any other odd $n=2\nu+1$
the result can be obtained by a differentiation \be\lb{receven}
G_{n+1}^{ret}\,=\,\frac{(-1)^{\nu-1}}{(2\pi)^{\nu-1}}
\frac{d^{\nu-1}}{(RdR)^{\nu-1}} \frac{\delta(T-R)}{R}, \ee so the
localization on the light cone is preserved. In particular, in
ten space-time dimensions one has \be\lb{9+1}
G_{9+1}^{ret}\,=\,\frac{1}{(2\pi)^3}\left(
\frac{\delta'''(T-R)}{R^4}+\frac{6\delta''(T-R)}{R^5}
+\frac{15\delta'(T-R)}{R^6}+\frac{15\delta(T-R)}{R^7}\right). \ee

For even $n=2\nu$ (odd dimension of space-time) the index of the
Bessel function is integer, and an integration in (\ref{sonin})
no more leads to the delta-function or its derivative. Using the
recurrence relation for $J_l(z)$: \be
J_{l+m}=-z^{l+m}\left(\frac{d}{zdz}\right)^m\left(z^{-l}
J_l\right), \ee one can obtain a relation similar to
(\ref{receven}) reducing the Green function of $n+1$ theory to
that in $2+1$ space-time: \be\lb{recodd}
G_{n+1}^{ret}\,=\,2\theta(T) \theta
(T^2-R^2)\frac{(-1)^{\nu-1}}{(2\pi)^{\nu-1}}
\frac{d^{\nu-1}}{(RdR)^{\nu-1}}\frac{1}{\sqrt{T^2-R^2}} , \ee the
lowest member of this family being \be G_{2+1}^{ret}=2\theta(T)
\theta (T^2-R^2)\frac{1}{\sqrt{T^2-R^2}}. \ee This function is
non-zero {\it inside} the future light cone. Therefore  a signal
from a short pulse source attains an observer at a distance $R$
after a time interval $R/c$, and then follows an infinitely long
tail. It is amusing to note how different is the retarded Green
function in eleven-dimensional space-time from that in ten
dimensions (\ref{9+1}): \be\lb{10+1}
G_{10+1}^{ret}\,=\,\frac{210}{(2\pi)^4}
\frac{\theta(T^2-R^2)}{(T^2-R^2)^{9/2}}. \ee

The occurrence of a tail in odd dimensions is by no means
surprising. Indeed, let us start with the  familiar case of four
dimensions representing the retarded Green function in the
following form \be G^{ret}_{3+1}=2\theta(T)\delta(T^2-R_3^2), \ee
where for simplicity we have assumed $x'^\mu=0$, so that
$R_3^2=x_1^2+x_2^2+x_3^2$. To pass to the Green function in $2+1$
dimensions it is sufficient to consider, instead of a
four-dimensional point-like source
$\delta^4(x)=\delta(t)\delta(x_1)\delta(x_2)\delta(x_3)$, a
line-like source $\delta^3(x)=\delta(t)\delta(x_1)\delta(x_2)$,
that is to integrate the previous source over $x_3$. Physically,
this means that waves coming to  the surface of a cylinder
$R_2^2=a^2$, where $R_2^2=x_1^2+x_2^2$, are collected not only
from directions normal to the axis $x_3$ (these show up exactly
after time $t=a$), but also from directions inclined to the axis
$x_3$. These oblique rays propagate longer times $t>a$, up to
$t=\infty$, and they constitute a tail. Mathematically, to find a
Green function in the next lower dimension one has to integrate
the initial Green function over the extra dimension. Integrating
the $3+1$-dimensional Green function (\ref{G4}) over $x_3$ one
obtains  \be \int G^{ret}_{3+1}
dx_3=\frac{2\theta(T)\theta(T^2-R_2^2)}{\sqrt{T^2-R_2^2}}, \ee
which is  the Green function $G^{ret}_{2+1}$.

One may wonder why a {\it double} integration over {\it two}
space-like dimensions leads again to the Green function localized
on the light cone and not developing a tail. The answer is simply
that tails are canceled by the second integration   due to
destructive interference. Let us start, e.g., from six dimensions,
representing the retarded Green function as \be
G^{ret}_{5+1}=-\frac{2}{\pi}\theta(T)\frac{d}{d\rho^2}
\delta(T^2-R_3^2-\rho^2), \ee where $\rho^2=x_4^2+x_5^2$. Now we
consider the    plane spanned by $x_4, x_5$ as a source and
integrate \be \int G^{ret}_{5+1} dx_4 dx_5=\int G^{ret}_{5+1}
2\pi \rho d\rho= 2\theta(T)\delta(T^2-R_3^2)=G^{ret}_{3+1}. \ee
The resulting Green function is again localized on the
(four-dimensional) light cone. Thus, the line source produces a
tail while the plane source does not. This explains
'periodicity'  of the Huygens property of the Green functions
with varying dimensions.
 %%%%%%%%%%%%%%%%%%%%%%%%%%%%%%%%%%%%%%%%%%%%%%%%%%%%%%%%%%%%%%%%%%%%%%%%%%%%%%%%%%%%%%%
\setcounter{equation}{0}
\section{2+1 theory}
In odd space-time dimensions, there is no much sense in splitting
the retarded Green functions into the self-force part and the
radiative part, since contributions from both of them are
non-local. Still, the mass renormalization can be consistently
performed. Near the coincidence limit the retarded $2+1$ Green
function diverges as \be G_{2+1}^{ret}\sim2
\frac{\theta(\sigma)}{|\sigma|}, \ee and therefore for $\sigma\to
0$ \be \frac{d}{d X^2} G_{2+1}^{ret}(X)\sim-
 \frac{\theta(\sigma)}{|\sigma|^3} \ee (the derivatives of the step
functions do not contribute to the integral over the world-line).
In view of the expansion (\ref{exp}), the singularity can be
completely removed by the renormalization of the mass term \be
m_0\du^\mu+e^2\int\;\frac{\du^\mu(s)}{|s-s'|}\;ds'=m\du^\mu. \ee
Now the divergence is logarithmic as could be expected for the
two-dimensional Coulomb potential. After this subtraction the
remaining integral over the world line becomes finite, and one
obtains the following equation  for the radiating charge in $2+1$
dimensions \be m\du^\mu(s)=e{F^\mu}_\nu u^\nu -
e^2\int_{-\infty}^s\;\left(\frac{2[X^\mu u^\nu(s)u_\nu(s')
-u^\mu(s')X^\nu u_\nu(s)]}{(X^2)^{3/2}}-
\frac{\du^\mu(s)}{|s-s'|}\right)\theta(X^2)\, ds'. \ee where
$X^\mu=X^\mu(s,s')=x^\mu(s)-x^\mu(s')$. The radiation reaction
force is given by the integral over an entire history of a charge.
%%%%%%%%%%%%%%%%%%%%%%%%%%%%%%%%%%%%%%%%%%%%%%%%%%%%%%%%%%%%%%%%%%%%%%%%%%%%%%%%%%%%%%%

\setcounter{equation}{0}
\section{$D>4$}
Let us start with even-dimensional space-times. Then the Green
functions contain (the derivatives of) $\delta(\sigma^2)$, so
only a finite number of Taylor expansions terms in $\sigma$ will
contribute. In dimensions $n+1$ with $n=2\nu+1$ we will have
instead of (\ref{sings}) \be \frac{d}{d X^2}
G^{self}(X)\sim\frac{d^\nu}{(d\sigma^2)^\nu}\delta(\sigma^2),\qquad
\frac{d}{d X^2} G^{rad}(X)\sim \frac{d^\nu}{(d\sigma^2)^\nu}\left(
\frac{\sigma}{|\sigma|}\delta(\sigma^2)\right). \ee Consequently,
one encounters the following integrals for the self-force \be
A^n_l=\int\;
\sigma^l\,\frac{d^{(n-1)/2}}{(d\sigma^2)^{(n-1)/2}}\delta(\sigma^2)d\sigma,
\ee and for the radiation reaction force \be B^n_l=\int\;
\sigma^l\,\frac{d^{(n-1)/2}}{(d\sigma^2)^{(n-1)/2}}\left(
\frac{\sigma}{|\sigma|}\delta(\sigma^2)\right) d\sigma, \ee with
$l\geq 2$. $A^n_l$ diverges for even $l$ satisfying $2\leq l<n$,
and vanishes for all other $l$. $B^n_l$ diverges for odd $l$
satisfying $3\leq l<n$, it has a finite value for $l=n$ and
vanishes for all other $l$.

To calculate the integral over the world-line now one has to
expand the quantity (\ref{exp}) up to the order $n$: \be\lb{expl}
4X^{[\mu}(s,s')u^{\nu]}(s')u_\nu(s)=\sum_{l=2}^n v_l\sigma^l
+O(\sigma^{n+1}), \ee where $v_l$ are polynomials of derivatives
of $u^\mu$ of total degree $l$: \be v_1=\du^\mu,\quad
v_2=-\frac23(\ddu^\mu+u^\mu\du^2 ),\quad v_3=\frac14
\frac{d^3u^\mu}{ds^3}+\frac16 \du^\mu \du^2+\frac34 u^\mu
(\du\cdot\ddu), \ee etc. To account properly for all powers of
$\sigma$ one also has to expand the argument $X^2$ of the
delta-functions up to higher orders in $\sigma$: \be
X^2=\sigma^2-\frac{1}{12}\du^2\sigma^4+\ldots   , \ee so that \be
\delta(X^2)=\delta(\sigma^2)-\frac{1}{12}\du^2\sigma^4
\delta'(\sigma^2) + \ldots\;. \ee From this analysis it is clear
that, with  dimension increasing by two, one acquires two more
 divergent integrals. Thus, already in six dimensions, we
will have three divergent integrals with different functional
dependence on dynamical variables, from which only one can be
removed by the mass renormalization. It is also worth noting that,
if one subtracts  the divergent integrals formally, the remaining
finite expression for the radiation reaction force would contain
$n$ derivatives of the four-velocity. It is unlikely that this
quantity could account indeed for the radiation reaction.

Now consider odd-dimensional space-times $n+1$ with $n=2\nu$. The
retarded Green function with both points on the world-line has
the leading singularity \be G^{ret}\sim
\frac{\theta(\sigma)}{|\sigma|^{2\nu-1}}. \ee To remove all
singularities from the integral over the world-line one has to
subtract $2\nu-1$ terms of an expansion of the integrand in
$\sigma$, from which only the lowest  subtraction   admits an
interpretation as a mass-renormalization. The situation is
exactly the same, as in even space-time dimensions, although
looking  somewhat differently.

%%%%%%%%%%%%%%%%%%%%%%%%%%%%%%%%%%%%%%%%%%%%%%%%%%%%%%%%%%%%%%%%%%%%%%%%%%%%%%%%%%%%%%%
\section{Conclusion}
To summarize: we have shown that the four-dimensional space-time
is distinguished as the unique one where a local differential
equation can be derived for a point charge moving in the external
electromagnetic field with account for the radiation reaction. In
$2+1$ dimensions the self-energy divergence can be removed by the
mass renormalization, but the resulting equation is
integro-differential, because of the tail effect in
odd-dimensional space-times. In dimensions higher than four,
divergences can no longer be removed by the renormalization of
parameters, so no consistent generalization of the Lorentz-Dirac
equation  exists for a point  charge. The analysis can be easily
extended to curved space-times and radiation fields of other
spins.

Clearly, the situation is not better in   quantum
electrodynamics: this theory is non-renormalizable   in
dimensions higher than four. But we have shown that even {\em
classical} theory of radiating point charges in higher dimensions
is not a fully consistent theory. It would be interesting to
investigate  classical renormalizability of  the equations of
motion of p-branes  interacting with fields of $p+1$ forms in
higher-dimensional space-times.
\section*{Acknowledgments}
This work was supported in part by the RFBR grant 00-02-16306.

\end{document}